\documentclass[prl, showpacs, twocolumn, superscriptaddress]{revtex4}
\usepackage{graphicx}
\usepackage{amssymb}
\usepackage{subfigure}

\begin{document}
\title{Parallel suppression of superconductivity and Fe moment in the collapsed tetragonal phase of Ca$_{0.67}$Sr$_{0.33}$Fe$_2$As$_2$ under pressure}

\author{J. R. Jeffries}
\affiliation{Condensed Matter and Materials Division, Lawrence Livermore National Laboratory, Livermore, CA 94550, USA}
\author{N. P. Butch}
\affiliation{Condensed Matter and Materials Division, Lawrence Livermore National Laboratory, Livermore, CA 94550, USA}
\affiliation{NIST Center for Neutron Research, National Institute of Standards and Technology, Gaithersburg, MD, 20899 USA}
\author{M. J. Lipp}
\affiliation{Condensed Matter and Materials Division, Lawrence Livermore National Laboratory, Livermore, CA 94550, USA}
\author{J. A. Bradley}
\affiliation{Condensed Matter and Materials Division, Lawrence Livermore National Laboratory, Livermore, CA 94550, USA}
\author{K. Kirshenbaum}
\affiliation{Center for Nanophysics and Advanced Materials, Department of Physics, University of Maryland, College Park, MD 20742, USA}
\author{S. R. Saha}
\affiliation{Center for Nanophysics and Advanced Materials, Department of Physics, University of Maryland, College Park, MD 20742, USA}
\author{J. Paglione}
\affiliation{Center for Nanophysics and Advanced Materials, Department of Physics, University of Maryland, College Park, MD 20742, USA}
\author{C. Kenney-Benson}
\affiliation{HP-CAT, Geophysical Laboratory, Carnegie Institute of Washington, Argonne, IL 60439, USA}
\author{Y. Xiao}
\affiliation{HP-CAT, Geophysical Laboratory, Carnegie Institute of Washington, Argonne, IL 60439, USA}
\author{P. Chow}
\affiliation{HP-CAT, Geophysical Laboratory, Carnegie Institute of Washington, Argonne, IL 60439, USA}
\author{W. J. Evans}
\affiliation{Condensed Matter and Materials Division, Lawrence Livermore National Laboratory, Livermore, CA 94550, USA}

\date\today

\begin{abstract}
Using non-resonant Fe $K{\beta}$ x-ray emission spectroscopy, we reveal that Sr-doping of CaFe$_2$As$_2$ decouples the Fe moment from the volume collapse transition, yielding a collapsed-tetragonal, paramagnetic normal state out of which superconductivity develops. X-ray diffraction measurements implicate the $c$-axis lattice parameter as the controlling criterion for the Fe moment, promoting a generic description for the appearance of pressure-induced superconductivity in the alkaline-earth-based 122 ferropnictides ({\it{A}}Fe$_2$As$_2$). The evolution of $T_c$ with pressure lends support to theories for superconductivity involving unconventional pairing mediated by magnetic fluctuations. 
\end{abstract}



\maketitle

Early in the research on iron pnictide superconductors, theoretical calculations strongly suggested that conventional, phonon-mediated superconductivity was incompatible with observed critical temperatures $T_c$ \cite{Boeri2008, Mazin2008, Yildrim2009, Hirschfeld2011}. Additionally, several systems have shown an empirical correlation between $T_c$ and the structural parameters \cite{Lee2008, Zhao2008, Kimber2009}. Since then, iron-based superconductors have proven a fertile playground for understanding how structural and magnetic degrees of freedom affect superconductivity \cite{Paglione2010, Johnston2010, Johannes2010, Stewart2011}. While there are no fewer than five systems sharing similar structural building blocks, the most widely studied ferropnictide superconductors are those that crystallize in the tetragonal ThCr$_2$Si$_2$ crystal structure. The so-named ``122'' compounds have the chemical formula $A$Fe$_2$$X_2$ (where $A$ is an alkaline-earth element, an alkali metal, or Eu and $X$ is a pnictogen atom), exhibit antiferromagnetic order at ambient pressure, and are amenable to a variety of chemical substitutions that suppress magnetism and induce superconductivity\cite{Paglione2010, Johnston2010, Stewart2011, Lumsden2010, Gooch2010, Jeevan2008}. 
 
With applied pressure or suitable chemical substitution, the parent 122 compounds undergo an isostructural volume collapse that is driven by the development of As-As bonding across the mirror plane of the crystal structure \cite{Goldman2009, Uhoya2011, Mittal2011, Jeffries2012, Saha2012}. This collapsed tetragonal (CT) phase abruptly cuts off the antiferromagnetic order and, in most cases, supports superconductivity in the vicinity of the truncated magnetic order \cite{Sefat2011}. The dissenting member of the 122 family, CaFe$_2$As$_2$, shows no signs of superconductivity in the CT phase when subjected to hydrostatic compression or isovalent P substitution, which mimics pressure through a reduction in the unit cell volume \cite{Yu2009, Kasahara2011}. This absence of superconductivity resonates with scenario where the Fe moments in the CT phase of CaFe$_2$As$_2$ are quenched \cite{Kreyssig2008, Soh2013}, supporting the strong link between the presence of magnetic fluctuations and the occurrence of superconductivity in this family of compounds \cite{Yildrim2009, Hirschfeld2011}. However, in Pr- and Nd-doped CaFe$_2$As$_2$, superconductivity can be induced from a normal state comprising non-magnetic Fe atoms \cite{Gretarsson2013}. These opposing behaviors with respect to the magnetic state of the Fe atoms question the roles of magnetic and charge fluctuations in inducing superconductivity in the 122 systems.
 
In order to tune the structural and magnetic degrees of freedom without charge doping, we have performed high-pressure experiments on Ca$_{0.67}$Sr$_{0.33}$Fe$_2$As$_2$, which provides a larger-volume system that decouples the quenching of Fe moments from the volume-collapse transition. Using non-resonant Fe $K{\beta}$ x-ray emission spectroscopy (XES) and x-ray diffraction (XRD), we report the evolution of the instantaneous Fe moments of Ca$_{0.67}$Sr$_{0.33}$Fe$_2$As$_2$ as a function of pressure and crystal structure. We find that the Fe moments persist into the CT phase, yielding a magnetic normal state out of which superconductivity develops. Consistent with theoretical calculations, the magnitude of the Fe moments appears to be largely controlled by the $c$-axis lattice parameter, providing a natural mechanism to explain the observed superconductivity in the 122 systems under pressure.

Single crystals of (Ca$_{0.67}$Sr$_{0.33}$)Fe$_2$As$_2$ were synthesized with a flux-growth technique previously described \cite{Saha2009}. For the XES measurements, small single crystals (11.3-keV data) or powder from the single crystal growth (20-keV data) were loaded into diamond anvil cells (DAC) that employed Be gaskets ($\sim$3~mm diameter). For XRD experiments, a powder was loaded into a spring steel gasket. The DAC was pressurized using a gas membrane, and the pressure in the sample chamber was calibrated using the shift of the ruby fluorescence line (XES data) or the lattice parameter of Cu (XRD data). Silicone oil (XES data) or neon (XRD data) were used as the pressure-transmitting media. 

The XES measurements were performed at beamline 16-IDD and the XRD measurements at 16-BMD of the High-Pressure Collaborative Access Team (HPCAT) at the Advanced Photon Source. For XES, 11.3- and 20.0-keV, micro-focused x-ray beams (approximately 35x50~$\mu$m$^2$) entered through one of the diamond anvils, while the emitted Fe $K{\beta}$ x-rays were collected after passing through the Be gasket. Fe $K{\beta}$ spectra were acquired by scanning a bent Si (440) analyzer in 0.25 eV steps; resolution was 1 eV. XRD data were collected in a transmission geometry using a 30 keV, 10x10~$\mu$m$^2$ micro-focused x-ray beam and a Mar345 Image plate. Analysis procedures for the XRD data are identical to Ref.~\onlinecite{Jeffries2012}. Additionally, for the 20.0 keV incident energy XES setup, x-ray diffraction was performed {\it{in situ}} to confirm the crystal structure of the Ca$_{0.67}$Sr$_{0.33}$Fe$_2$As$_2$ specimen (see Supplementary Materials). Low-temperature XES and XRD measurements were performed in a He-flow cryostat.

%%%%%%%%%%%%%%%%%%%%%%%%%%%%%%%%%%%
\begin{figure}[t]
%COMMENT LINE h=here, t=top, b=bottom, p=separate figure page
\begin{center}\leavevmode
\includegraphics[scale=0.31]{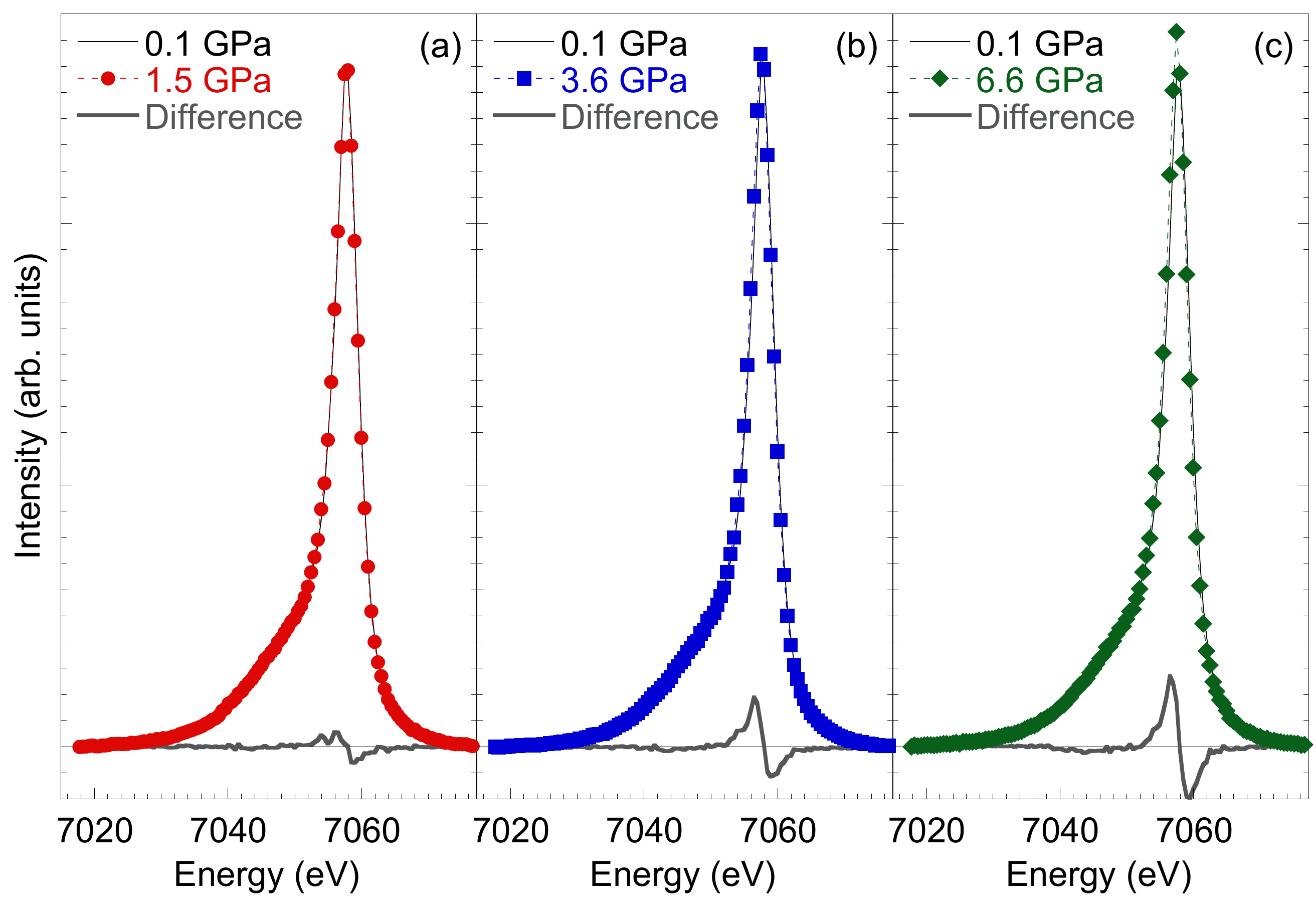}
\caption{(color online) Room-temperature, unit-normalized XES spectra of (Ca$_{0.67}$Sr$_{0.33}$)Fe$_2$As$_2$ under pressure (dashed lines, closed symbols) at (a) 1.5, (b) 3.6, and (c) 6.6 GPa as compared with the spectrum at 0.1 GPa (black line, no symbols). The differences between the XES data at each pressure and those at 0.1 GPa are shown as the grey solid lines below the data. Measurement errors due to counting statistics are less than 1\%.}\label{XES_Diff}
\end{center}
\end{figure}
%%%%%%%%%%%%%%%%%%%%%%%%%%%%%%%%%%%

The Fe $K{\beta}$ spectra are sensitive, bulk probes of the instantaneous Fe moment due to atomic $3p$-$3d$ orbital overlap. An incoming x-ray with an energy greater than the Fe K-edge can excite a $1s$ electron from the core of an Fe atom into the continuum, leaving a $1s$ core-hole that can be filled with a decay from the $3p$ shell. This $3p-1s$ decay in Fe emits an x-ray, the $K{\beta}$ line, at 7058 eV and leaves a final-state core-hole in the $3p$ level. In the presence of a $3d$ moment, the spin degeneracy of this $3p$ core-hole is lifted, producing an energy difference between the spin states of the $3p$ core-hole, and yielding an emission spectrum with a predominant $K{\beta}$ peak and a weaker satellite $K{\beta}^{\prime}$ peak \cite{Lin2005, Rueff2010}. For small-moment systems, like Ca$_{0.67}$Sr$_{0.33}$Fe$_2$As$_2$, the $K{\beta}^{\prime}$ portion of a spectrum appears as a shoulder, rather than a distinct peak, on the low-energy side of the main $K{\beta}$ peak. Unfortunately, a low-energy shoulder exists even in non-magnetic Fe systems \cite{Kumar2011}, and, as such, a systematic analysis of the {\it{entire}} emission spectrum is necessary to extract a quantitative Fe moment. Here we follow the integrated absolute difference (IAD) method to evaluate the Fe moment in Ca$_{0.67}$Sr$_{0.33}$Fe$_2$As$_2$ under pressure \cite{Vanko2006}. An IAD analysis proceeds as follows: each spectrum is normalized such that its integral is unity, a unit-normalized reference spectrum is subtracted from each spectra to yield a difference spectrum, and the absolute values of the difference spectra are integrated to yield quantitative IAD values that are proportional to the Fe moment for each spectrum. For tetrahedrally coordinated Fe atoms, the IAD value can be converted to an Fe moment (in ${\mu}_B$) using the calibration provided by Gretarsson, {\it{et al}} \cite{Gretarsson2011}.  Here, we use our room-temperature, 0.1-GPa data as the reference; Fe moment measurements are thus measured relative to that at 0.1 GPa, which is near enough to ambient pressure to use the ambient-pressure value of about 1.1 $\mu_B$ for the $A$Fe$_2$As$_2$ systems\cite{Gretarsson2011}.

Example room-temperature XES measurements are shown in Figures~\ref{XES_Diff}a-c, which display the unit-normalized Fe $K{\beta}$ spectra at 1.5, 3.6, and 6.6 GPa as compared to the spectrum at 0.1 GPa. The shoulder associated with the $K{\beta}^{\prime}$ satellite is clearly evident in the low-energy asymmetry of the main peak. The differences between the reference spectrum (0.1-GPa data) and the high-pressure spectra are shown as the solid, grey lines. With increasing pressure, the difference spectra clearly grow larger. The negative dip seen at 6.6 GPa near 7045 eV indicates a loss of spectral weight in the $K{\beta}^{\prime}$ satellite portion of the spectrum, implying a reduction in the Fe moment with increasing pressure. 

\begin{figure}[t]
%COMMENT LINE h=here, t=top, b=bottom, p=separate figure page
\begin{center}\leavevmode
\includegraphics[scale=0.47]{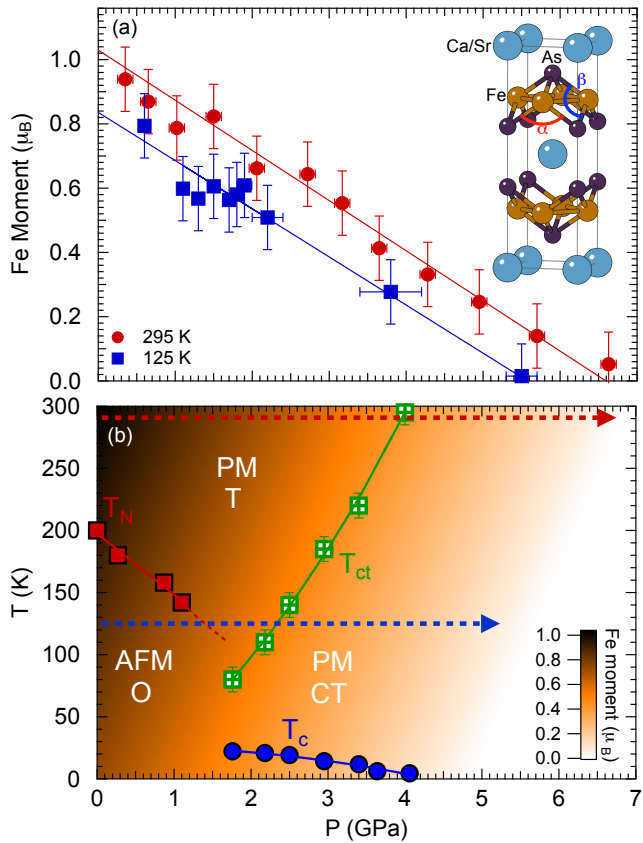}
\caption{(color online) (a) The pressure-dependent Fe moment of (Ca$_{0.67}$Sr$_{0.33}$)Fe$_2$As$_2$ at 295 and 125 K; solid lines through the data are linear fits. Horizontal error bars are the larger of 0.1 GPa or the difference between the pressures before and after collection of the spectra. Vertical error bars are 0.1 $\mu_B$ estimated from the noise in the spectra. The inset displays the crystal structure of (Ca$_{0.67}$Sr$_{0.33}$)Fe$_2$As$_2$, and labels the two As-Fe-As bond angles. (b) The electronic and structural phase diagram of (Ca$_{0.67}$Sr$_{0.33}$)Fe$_2$As$_2$, including a contour plot of the Fe moment in $\mu_B$. Data points and error bars are described in Ref. \onlinecite{Jeffries2012}. The arrows indicate the approximate paths of the XES measurements.}\label{Moment}
\end{center}
\end{figure}

The IAD values from each pressure at 300 and 125 K have been converted to Fe moments, and these values are shown in Fig.~\ref{Moment}a. At both 125 and 300 K, pressure drives the Fe moment down linearly; the slope of the pressure dependence of the Fe moment, -0.15 ${\mu}_B$/GPa, is identical within our experimental precision for both temperatures. The measured isotherms are overlaid on the phase diagram shown in Fig.~\ref{Moment}b. At 300 K, the structure of (Ca$_{0.67}$Sr$_{0.33}$)Fe$_2$As$_2$ undergoes a tetragonal (T) to collapsed-tetragonal (CT) phase transition near 4 GPa \cite{Jeffries2012}. At 125 K and ambient pressure, (Ca$_{0.67}$Sr$_{0.33}$)Fe$_2$As$_2$ is antiferromagnetically ordered (AFM), and the crystal structure is orthorhombic (O) rather than tetragonal. With pressure at 125 K, the system undergoes an O-T transformation that destroys AFM order in favor of a paramagnetic (PM) state at about 2 GPa; the T-CT phase transition occurs near 2.8 GPa. The Fe moment evolves relatively smoothly through each of these structural/magnetic phase transitions.

%%%%%%%%%%%%%%%%%%%%%%%%%%%%%%%%%%%
%%%%%%%%%%%%%%%%%%%%%%%%%%%%%%%%%%%

In addition to the phase boundaries of Fig.~\ref{Moment}b, a contour plot of the Fe moment in T-P space is included. A linear interpolation is employed between 300 and 125 K, and the same trend is extrapolated to lower temperatures. This linear extrapolation is consistent with T-dependent, ambient-pressure XES measurements (see Supplementary Materials) as well as the nearly linear temperature dependence (for $T{\gtrsim}$ 50 K) of the moment and lattice parameters of (Ca$_{0.78}$La$_{0.22}$)Fe$_2$As$_2$ \cite{Gretarsson2013}. At low pressures and intermediate temperatures, the CT phase abruptly destroys the ordered AFM state, but the Fe moment evolves continuously through this transition. This persistence of the Fe moment into the CT phase implies that the destruction of AFM order is not precipitated by the quenching of the local moments. Rather, a more favorable scenario would likely be one where the CT phase supports enhanced spin fluctuations or alters coupling strengths that lead to the destruction of magnetic order \cite{Johannes2010}

At 1.8 GPa, the extrapolated moment is about 0.4 $\mu_B$ at 25 K, just above the value of $T_c{\approx}$22 K. With increasing pressure, the linear suppression of the Fe moment is mirrored by the nearly linear suppression of superconductivity, with $T_c$ disappearing within about 0.5 GPa of the pressure at which the Fe moment is expected to quench. These observations suggest that the superconducting state of  (Ca$_{0.67}$Sr$_{0.33}$)Fe$_2$As$_2$ not only develops out of a magnetic normal state, but also that $T_c$ is closely coupled to the magnitude of the Fe moment. This persistent moment near $T_c$ is comparable to (Ca$_{0.78}$La$_{0.22}$)Fe$_2$As$_2$, where an approximately 0.5 $\mu_B$ moment is present with $T_c{\approx}$35 K \cite{Gretarsson2013, Saha2012}. The correlation between Fe moment and $T_c$ is consistent with theories that evoke spin fluctuations as the pairing mechanism for superconductivity \cite{Hirschfeld2011}.

%%%%%%%%%%%%%%%%%%%%%%%%%%%%%%%%%%%
\begin{figure}[t]
%COMMENT LINE h=here, t=top, b=bottom, p=separate figure page
\begin{center}\leavevmode
\includegraphics[scale=0.42]{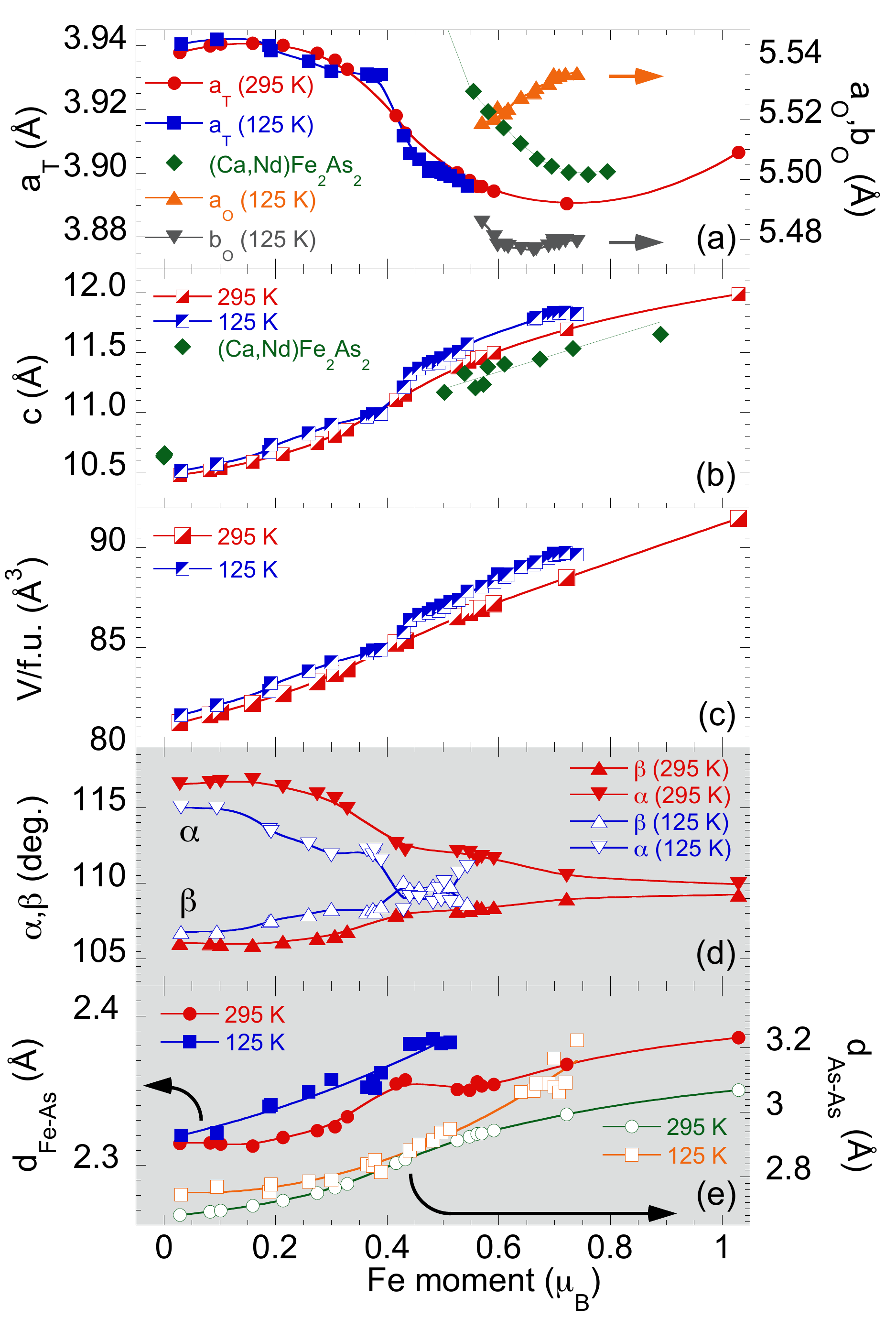}
\caption{(color online) Crystal structure parameters of (Ca$_{0.67}$Sr$_{0.33}$)Fe$_2$As$_2$ as a function of Fe moment for different temperatures: (a) $a_T$, $a_O$, and $b_O$, lattice parameters; (b) $c$ lattice parameter; and (c) unit cell volume per formula unit $V$/f.u. The internal degrees of freedom of the crystal structure: (d) As-Fe-As bond angles (see inset of Fig.~\ref{Moment}), and (e) the Fe-As and As-As bond distances. The $a$- and $c$-axis lattice parameters (from temperature-dependent measurements) for (Ca$_{0.92}$Nd$_{0.08}$)Fe$_2$As$_2$ are included in (a) and (b), respectively \cite{Gretarsson2013}. Error bars are smaller than the data markers.}\label{Structure}
\end{center}
\end{figure}
%%%%%%%%%%%%%%%%%%%%%%%%%%%%%%%%%%%

Unlike the temperature dependent moments of Pr- and Nd-doped CaFe$_2$As$_2$ \cite{Gretarsson2013}, which discontinuously quench upon entering the CT phase, the Fe moments at 300 and 125 K (Fig.~\ref{Moment}a) of (Ca$_{0.67}$Sr$_{0.33}$)Fe$_2$As$_2$ do not exhibit any large discontinuities upon crossing the T-CT or O-T phase boundaries. While this may at first appear inconsistent, a closer examination of the effect of structure on the Fe moments reveals common behavior. Figure~\ref{Structure} shows the evolutions of the following structural parameters of (Ca$_{0.67}$Sr$_{0.33}$)Fe$_2$As$_2$ at 300 and 125 K as functions of the Fe moment: the orthorhombic ($a_O$ and $b_O$) and tetragonal ($a_T$) basal plane lattice parameters; the $c$-axis lattice parameter; the volume per formula unit ($V/f.u.$); the As-Fe-As bond angles; as well as the Fe-As bond length ($d_{Fe-As}$) and the As-As mirror plane spacing ($d_{As-As}$). The $a_T$, $c$, and $V$/f.u. parameters show significant overlap at 300 and 125 K, whereas the Fe-As separation and the As-Fe-As bond angles do not. This implies that the lattice parameters, rather than the internal structural degrees of freedom, are the dominant factors controlling the magnitude of the Fe moment in (Ca$_{0.67}$Sr$_{0.33}$)Fe$_2$As$_2$. This correlation is consistent with the supposition that structure plays an important role in setting the magnetic energy scales of the $A$Fe$_2$As$_2$ systems \cite{Kirshenbaum2012}.

Including data from (Ca$_{0.92}$Nd$_{0.08}$)Fe$_2$As$_2$ further illuminates the correlations between structure and Fe moment.  Figs.~\ref{Structure}(a) and (b) include structure data for (Ca$_{0.92}$Nd$_{0.08}$)Fe$_2$As$_2$ taken as a function of temperature (note: (Ca$_{0.78}$La$_{0.22}$)Fe$_2$As$_2$ behaves nearly identical) \cite{Gretarsson2013, Saha2012}. While the T-dependent (and thus Fe-moment-dependent) $a$-axis lattice parameter of Nd-doped CaFe$_2$As$_2$ is distinctly different from that of (Ca$_{0.67}$Sr$_{0.33}$)Fe$_2$As$_2$ under pressure, the $c$-axis lattice parameters for these compounds follow strikingly similar behavior as a function of Fe moment. The temperature-induced T-CT transition in (Ca$_{0.92}$Nd$_{0.08}$)Fe$_2$As$_2$ results in a large ($\sim$5\%) change in the $c$-axis lattice parameter that yields a zero-moment CT phase with $c<$10.65~\AA{} \cite{Gretarsson2013}. When the $c$-axis lattice parameter of (Ca$_{0.67}$Sr$_{0.33}$)Fe$_2$As$_2$ is compressed to the same value $c<$10.65~\AA{} at either 300 or 125 K, the Fe moment is similarly lost. The strong correlation between $c$-axis lattice parameter and the Fe moment is in excellent qualitative agreement with theoretical calculations \cite{Yildrim2009, Johannes2010}, but it should be noted that the pressure-dependent volume contraction of (Ca$_{0.67}$Sr$_{0.33}$)Fe$_2$As$_2$ is almost entirely driven by the $c$-axis lattice parameter (the $a$-axis lattice parameter is actually expanding over a portion of phase space). Therefore, whether the mechanism for moment loss is driven by changes in the atomic volume of Fe, as suggested for mantle minerals \cite{Speziale2005}, or some $c$-axis-dependent internal coordination remains ambiguous.

These experimental results challenge the notion that the Fe moment is universally quenched in the CT phase of the 122 systems, and instead promote a generic description of the 122 systems under pressure. With increasing pressure, As-As bonding develops across the mirror plane of the crystal structure, isostructurally collapsing the structure and truncating the AFM state. While the Fe moments in the CT phase of undoped CaFe$_2$As$_2$ are quenched \cite{Kreyssig2008, Soh2013}, we have shown that the CT phase can in fact support a substantial Fe moment ($\sim$0.4 $\mu_B$) that appears to be strongly coupled to the $c$-axis lattice parameter, which is controlled by the size of the alkaline earth atom, thermal contraction, and the volume change induced by the CT phase. Details of the crystal structure thus govern the magnitude of the Fe moment in the CT phase, potentially providing a paramagnetic normal state out of which superconductivity can develop even in the absence of charge doping. That the pressure-dependent suppression of $T_c$ tracks that of the Fe moment supports the picture of an unconventional superconducting state mediated by magnetic fluctuations. This generic picture suggests that driving the CT phase to ambient pressure while maintaining an Fe moment should be a promising route to high-temperature, magnetically mediated superconductivity in the 122 systems.

We are grateful to K. Visbeck for assistance with cell preparations. Portions of this work were performed under LDRD (14-ERD-041). Lawrence Livermore National Laboratory is operated by Lawrence Livermore National Security, LLC, for the U.S. Department of Energy, National Nuclear Security Administration under Contract DE-AC52-07NA27344. Portions of this work were performed at HPCAT (Sector 16), Advanced Photon Source (APS), Argonne National Laboratory. HPCAT operations are supported by DOE-NNSA under Award No. DE-NA0001974 and DOE-BES under Award No. DE-FG02-99ER45775, with partial instrumentation funding by NSF. APS is supported by DOE-BES, under Contract No. DE-AC02-06CH11357. Beamtime was provided through the Carnegie-DOE Alliance Center (CDAC) and the APS General User Program (GUP). This work was partially supported by AFOSR-MURI Grant No. FA9550-09-1-0603.

\end{document}